\documentclass[aps,preprint,superscriptaddress,nofootinbib]{revtex4}

\usepackage[usenames]{color}
\usepackage{graphicx}
\usepackage{amsmath}
\usepackage{amsfonts}
\usepackage{amssymb}
\usepackage{mathrsfs}
\usepackage{bm}
\usepackage{verbatim}
\usepackage{cancel}

\newcommand{\beq}{\begin{equation}}
\newcommand{\eeq}{\end{equation}}
\newcommand{\bea}{\begin{eqnarray}}
\newcommand{\eea}{\end{eqnarray}}

\newcommand{\exctL}{\Psi}

\preprint{CTP-SCU/2015011}

\begin{document}

\title{Chiral dynamics of $S$-wave baryon resonances}

\author{Bingwei Long}
\email{bingwei@scu.edu.cn}
\affiliation{Center for Theoretical Physics, Department of Physics, Sichuan University, 29 Wang-Jiang Road, Chengdu, 610064, China}

\date{July 22, 2016}

\begin{abstract}

As the pion mass approaches a critical value $m_\pi^\star$ from below, an $S$-wave resonance crosses the pion-baryon threshold and becomes a bound state with arbitrarily small binding energy, thus driving the scattering length to diverge. I explore the consequences of chiral symmetry for the values of $m_\pi$ close to $m_\pi^\star$. It turns out that chiral symmetry is crucial for an $S$-wave resonance to be able to stand very near the threshold and in the meantime to remain narrow, provided that the mass splitting is reasonably small. The effective range of pion-baryon scattering is unexpectedly large, proportional to $ 4\pi f_\pi^2/m_\pi^3$ when $m_\pi$ is around $m_\pi^\star$. As a result, this unexpected large length scale causes universality relations to break down much sooner than naively expected.
\end{abstract}

\maketitle

From the viewpoint of nonrelativistic potential models, $S$-wave hadronic resonances are less common than higher-wave ($l > 0$) resonances because $S$ waves do not have centrifugal barriers to prevent a component particle from leaving the other. Therefore, hadronic $S$-wave resonances possess special merits because they are less amenable to simple model potentials, and hence understanding them gives a glimpse of the underlying theory---quantum chromodynamics (QCD). If the $S$-wave resonance arises from an excited baryon coupled to a low-lying baryon and a pion, chiral perturbation theory (ChPT), which is a low-energy effective field theory of QCD, can be a viable tool to investigate the system, for pion-baryon interactions are restricted by chiral symmetry.

The present manuscript is concerned with even more special cases where the $S$-wave resonance is near the pion-baryon threshold. Near-threshold higher-wave ($l > 0$) resonances are less surprising because it can be quite generally shown that a single-parameter fine-tuning of the two-body interaction gets the resonance arbitrarily close to threshold, eventually turning it into a bound state~\cite{Taylor-Scattering}. But it usually takes a two-parameter fine-tuning to move an $S$-wave resonance close to threshold, as indicated by the scattering length and effective range~\cite{Hyodo:2013iga} in the effective range expansion or by two independent parameters of a model potential~\cite{Hanhart:2014ssa}. I show that with the assistance of chiral symmetry, such extreme fine-tunings are no longer needed for an $S$-wave resonance to be near pion-baryon threshold.

Remarkably, there is a real-world example of near-threshold $S$-wave baryon resonance: charmed baryon $\Lambda_c^+(2595)$ coupled to the $\pi \Sigma_c(2455)$ channel, with $\Delta \simeq m_\pi^\text{phy} = 138\, \text{MeV}$, where $\Delta$ is the $\Lambda_c^+(2595) - \Sigma_c(2455)$ mass splitting. Therefore, $\Lambda_c^+(2595)$ is considered as the example in the manuscript. Other ChPT-based, phenomenological, or lattice investigations on charmed baryons can be found in Refs.~\cite{Lutz:2003jw, GarciaRecio:2006wb, GarciaRecio:2008dp, Romanets:2012ce, Lutz:2014jja, Lu:2014ina, Namekawa:2013vu, Brown:2014ena, Bali:2015lka} and references therein.

But the relevance of near-threshold $S$-wave resonances is not necessarily confined to $\Lambda_c^+(2595)$. With variable quark masses, lattice QCD can create other hadronic worlds, labeled by various values of the pion mass $m_\pi$. At low energies, the ``otherworldly'' nuclear physics has started to attract interest~\cite{Beane:2013br, Barnea:2013uqa}. Near-threshold $S$-wave resonances may emerge when $m_\pi$ approaches a certain critical value $m_\pi^\star$, at which the scattering length $a$ diverges.

Such emergences may be seemingly inevitable because $\Delta$ as a nonchiral quantity is expected to vary slower than $m_\pi$, so the resonance should become stable when $m_\pi$ crosses $\Delta$ from below. Therefore, $m_\pi^\star$ is just $\Delta$ plus subleading corrections: $m_\pi^\star \simeq \Delta$. $a$ diverges because the binding energy can be made arbitrarily small by tuning $m_\pi - \Delta$.

However, the above simple argument applies only to higher-wave resonances. In $S$ waves, the transition from a resonance to a bound state is more complicated. The general theory of two-body scattering tells us that the $S$-wave resonance normally broadens out near threshold and turns into a pair of virtual states before becoming stable (see, for example, Refs.~\cite{Hyodo:2013iga, Hanhart:2014ssa}). In the complex plane of the magnitude of the center-of-mass (CM) momentum $k$, the transition is illustrated as the pair of resonance poles coalescing on the lower half imaginary axis before splitting into two virtual poles, and then heading respectively upward and downward. The virtual pole on the top eventually moves above the real axis and becomes a bound state pole. For the $S$-wave resonance to remain narrow right before it turns virtual, the coalescing point must be rather close to threshold, implying the existence of another infrared mass scale in addition to the vanishing $m_\pi - \Delta$, which echoes the aforementioned two-parameter fine-tuning.

It is the main message of the present paper that thanks to chiral symmetry, the extra fine-tuning is waived for near-threshold $S$-wave pion-baryon resonances, as long as $\Delta \ll \sqrt{4\pi} f_\pi \simeq 328$ MeV. The mechanism is that chiral symmetry requires that a rather large (but not diverging) value be taken by the effective range $r$ by constraining how the excited baryon is coupled in the $S$ wave to the pion-baryon continuum. For $m_\pi$ close to $m_\pi^\star \simeq \Delta$, the leading-order (LO) value of $r$ is found to be
\begin{equation}
  r = -\frac{4\pi f_\pi^2}{h^2 m_\pi^3} \, , \label{eqn:rlo}
\end{equation}
where the pion decay constant $f_\pi = 92.4$ MeV; $h$ is the dimensionless coupling constant of the resonance to the pion-baryon system and it is assumed to be of the order of unity.

But there is still one fine-tuning remaining: $m_\pi$ close to $m_\pi^\star$, $(m_\pi^\star - m_\pi) \to 0$. A second motivation for this line of research is to study how this fine-tuning is propagated through hadronic systems. (A precedent of such investigations is the real-world nucleon-nucleon system, whose large values of scattering lengths are suspected to result from the physical pion mass being in close proximity to a critical value~\cite{Beane:2006mx}.) In the immediate neighborhood of $m_\pi^\star$, many dimensionful quantities scale only with $(m_\pi - m_\pi^\star)$, a rule known as universality~\cite{Braaten:2004rn}. However, the simultaneous emergence of two large length scales, $a$ and $r$, by a single fine-tuning $(m_\pi^\star - m_\pi) \to 0$ invalidates the universality relations that account for only the large value of $a$. I use the binding energy to demonstrate how $r$ affects the threshold physics.

Two-flavor chiral symmetry suffices to demonstrate the points I make. Regardless of the isospin of the $S$-wave resonance, the lowest-order coupling of the resonance to the pion and baryon must involve one time derivative on the pion field. Ensured by chiral symmetry and parity conservation, this is the single most important feature of an $S$-wave baryon resonance, and it is the foundation of what is developed here. The heavy-baryon Lagrangian terms with Weinberg's chiral index \cite{Weinberg:1978kz, Jenkins:1990jv} $\nu = 0 $ are
\begin{equation}
    \begin{split}
    \mathcal{L}^{(0)} &= {\Sigma^a}^\dagger \left[i\partial_0 \delta_{ab} + \frac{i}{f_\pi^2} \left( \pi^a \dot{\pi}^b - \pi^b \dot{\pi}^a \right) \right] \Sigma^b \\
    &\quad + \exctL^\dagger \left( i\partial_0 - \Delta \right) \exctL + i\frac{g_\Sigma}{f_\pi} \epsilon_{abc} {\Sigma^a}^\dagger \vec{\sigma} \cdot \vec{\nabla} \pi^b \Sigma^c \\
    & \quad +  \frac{h}{\sqrt{3}f_\pi} \left( {\Sigma^a}^\dagger \dot{\pi}^a \exctL + h.c. \right)  + \cdots
    \label{eqn_nu0}
    \end{split}
\end{equation}
Here $\exctL$ ($\Sigma$) is the field that annihilates $\Lambda_c^+(2595)$ [$\Sigma_c(2455)$] and $g_\Sigma$ the axial coupling of $\Sigma_c(2455)$. \footnote{The $D$-meson-nucleon system can be integrated out here because $DN$ has to be quite off shell to be relevant, with the CM momentum around $\sqrt{2 \mu \Delta_{DN}} \simeq 510$ MeV, where $\mu$ is the reduced mass of $DN$ and $\Delta_{DN}$ is the CM energy difference between $\Lambda_c^+(2595)$ and the $DN$ threshold.}

Now we turn to construction of the $S$-wave amplitude for $\pi \Sigma_c$ elastic scattering. When $m_\pi$ is near $m_\pi^\star$, either below or above, $\Lambda_c^+(2595)$ remains a near-threshold phenomenon and the pion is nonrelativistic. Therefore, $k$ and the energy shift of the resonance from threshold $\delta \equiv \Delta - m_\pi$ are both much smaller than $m_\pi$: $k/m_\pi \ll 1$ and  $|\delta| / m_\pi \ll 1$. The recoil effects of the pion are systematically included, whereas those of the baryon are not considered here, due to its much larger mass.

While more formal treatments of heavy pions can be found in Refs.~\cite{Beane:2002aw, Fleming:2007rp, Braaten:2015tga}, I choose to use the usual ChPT framework in which the pions are created and/or annihilated by a relativistic field.

With the incoming (outgoing) 4-momentum of $\pi$ denoted by $k_\mu$ ($k'_\mu$) and that of $\Sigma_c$ by $p_\mu$ ($p'_\mu$), I write the isoscalar $S$-wave $\pi \Sigma_c$ potentials as the following two pieces:
The $s$-channel exchange of $\exctL$ is
\begin{equation}
  v_s = \frac{h^2}{f_\pi^2} \frac{k_0 k_0'}{k_0 + p_0 - \Delta} = \frac{h^2 m_\pi^2}{f_\pi^2(E - \delta)} \left[ 1 + \mathcal{O}\left(\frac{Q^2}{m_\pi^2}\right)  \right]\, , \label{eqn_vs}
\end{equation}
where $E$ is the CM energy, and the Weinberg-Tomozawa (WT) term
\begin{equation}
  v_{\text{WT}} = \frac{3 (k_0 + k_0')}{2 f_\pi^2} = \frac{3 m_\pi}{f_\pi^2} \left[ 1 + \mathcal{O}\left(\frac{Q^2}{m_\pi^2}\right)  \right] \label{eqn_vWT} \, .
\end{equation}
The $u$-channel exchanges of $\Sigma$ or $\exctL$ are not considered because they involve two powers of $Q$, and are thus smaller than $v_\text{WT}$ by $\mathcal{O}(Q^2/m_\pi^2)$, where $Q$ denotes generically external momenta.

\begin{figure}
        \includegraphics[scale=0.4]{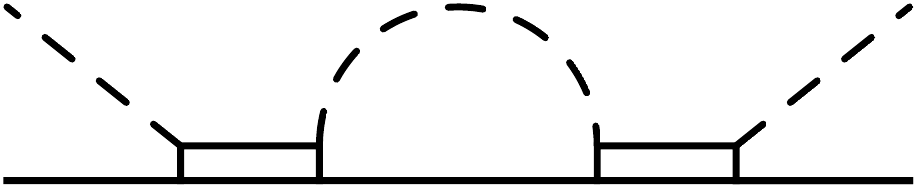}
    \caption{Once iterated $s$-channel exchange of $\exctL$ in $\pi \Sigma_c$ scattering. The solid, dashed, and double lines represent $\Sigma_c$, $\pi$, and $\exctL$, respectively.}
    \label{fig:pisigmac}
\end{figure}

Resummation of $v_s$ gives rise to the desired nonperturbative physics, but an argument for its necessity in the power-counting language helps us understand theoretical uncertainties of the EFT-based conclusions~\cite{Bedaque:2003wa, Pascalutsa:2002pi, Long:2009wq}. Figure~\ref{fig:pisigmac} shows two insertions of $v_s$, connected by a pion-baryon loop. When $E-\delta$ in the denominator of $v_s$ is as small as the $\Psi$ self-energy, all diagrams with serial insertions of $v_s$ are equally important, hence the resummation.

Let us first power count the nonrelativistic pion-baryon loop, shown as part of Fig.~\ref{fig:pisigmac}. The fact that the pion is nonrelativistic modifies in several respects the standard power counting~\cite{Weinberg:1978kz}. The 3-momentum of the pion internal line is of $Q$ and the energy $m_\pi + Q^2/m_\pi$; therefore, the pion propagator is counted as $1/Q^2$. The baryon propagator is static, and the energy flowing through it is of the same order as the kinetic energy of the pion. So, the baryon propagator is counted as $m_\pi/Q^2$. With the internal pion 4-momentum denoted by $l$, the integration volume $\int d^4l$ contributes a factor $\sim Q^5/m_\pi$, in which $\int dl_0 \sim Q^2/m_\pi$ and $\int d^3l \sim Q^3$. In addition, the numerical factor coming out of a nonrelativistic loop is normally $1/4\pi$, compared with that of a relativistic loop, $1/16\pi^2$. In conclusion, a nonrelativistic pion-baryon loop contributes a factor of $Q/4\pi$.

Together with the coupling of $\Psi$ to $\pi \Sigma_c$, the LO self-energy of $\Psi$ will be $\sim m_\pi^2 Q/(\sqrt{4\pi} f_\pi)^2$, in contrast with $\sim Q^3/(4\pi f_\pi)^2$ in the case of a relativistic pion. The appearance of the intermediate mass scale $\sqrt{4\pi} f_\pi = 328$ MeV is worth taking note of. It suggests that even though $Q$ is small for nonrelativistic pions, the self energy of $\Psi$ is less suppressed than expected from the standard ChPT counting.

The criterion for resummation is when $v_s$ and a pion-baryon loop combine to contribute a factor of $\mathcal{O}(1)$,
\begin{equation}
  \frac{m_\pi^2}{f_\pi^2 (E - \delta)} \frac{Q}{4\pi} = \frac{\epsilon Q}{E - \delta} \gtrsim 1 \, ,\label{eqn:delta_ine}
\end{equation}
where $\epsilon$ is a function of $m_\pi$ and it is by our choice a small parameter,
\begin{equation}
  \epsilon(m_\pi) \equiv \left(\frac{m_\pi}{\sqrt{4\pi} f_\pi}\right)^2 \, .
\end{equation}
This defines the so-called resonance region, a kinematic window in terms of $E$ inside of which it is necessary to dress the bare $\Psi$ propagator. In the resonance region, $v_\text{WT}$ is smaller than $v_s$ by $\mathcal{O}(\epsilon Q/m_\pi)$. In addition, $v_\text{WT}$ in itself is perturbative in the sense that two insertions of $v_\text{WT}$ bring suppression to the tree-level $v_\text{WT}$ by $\mathcal{O}(\epsilon Q/m_\pi)$.

It is straightforward to sum up the geometric series of connected $v_s$'s. Not surprisingly, the LO amplitude $T^{(0)}$ has the form of the effective range expansion,
\begin{equation}
    \begin{split}
        T^{(0)}(k) = \frac{4\pi}{-\gamma_0 + \frac{r_0}{2}k^2 - ik}\, ,
    \end{split}
\end{equation}
with
\begin{equation}
    \gamma_0 = - \frac{\delta}{h^2 \epsilon} \, , \quad
    r_0 = - \frac{4\pi f_\pi^2}{h^2 m_\pi^3} = -(h^2 \epsilon m_\pi)^{-1} \, , \label{eqn:a0r0}
\end{equation}
where the subscripts 0 on $\gamma_0$ and $r_0$ indicate that the inverse scattering length and effective range are LO values.
Since $m_\pi^\star$ is defined to be the value of $m_\pi$ where the scattering length diverges, one finds at LO $\delta = m_\pi^\star - m_\pi$. If $\Delta$ is reasonably small $\Delta \ll \sqrt{4\pi} f_\pi$, the effective range gains a rather large value, provided that $h$ is naturally sized. In fact, by assuming $\pi \Sigma_c$ to be the dominant decay channel Ref.~\cite{Hyodo:2013iga} determines from the PDG values for mass and width of $\Lambda^+_c(2595)$ that $r = 19.5$ fm, which gives $h = 0.64$, consistent with the assumption that $h$ is of the order of unity.

Even when the threshold is outside the resonance region, we may still use Eq.~\eqref{eqn:a0r0} to determine $\gamma_0$ and $r_0$, under the condition that the WT term remains the subleading contribution to the threshold scattering. To this end, $v_\text{WT}$ must be much smaller than $v_s$ for $k \simeq 0$, which sets the validity range for Eq.~\eqref{eqn:a0r0} $\delta \ll m_\pi$, a condition that we have already presumed. Note that $\delta/m_\pi$ and $\epsilon$ are independent small parameters, so the results and the discussion do not rely on whether they are correlated.

The LO amplitude has two poles on the $k$ plane:
\begin{equation}
  k_{\pm} = - h^2 \epsilon m_\pi \left(i \pm \sqrt{\frac{2\delta}{h^4 \epsilon^2 m_\pi} - 1} \right) \, .\label{eqn:kpm}
\end{equation}
The poles move as $\delta$ becomes negative from positive. Starting as two conjugate resonance poles in the lower half plane, they move toward each other and coalesce on the lower half imaginary axis when $\delta = \frac{1}{2}h^4 \epsilon^2 m_\pi$. The imaginary part of the resonance pole position $k_I = - h^2 \epsilon m_\pi$ (or the inverse effective range $1/r$) is small and changes slowly with respect to $\delta$. This helps the resonance remain narrow when it is located very near threshold, until $\delta \sim \frac{1}{2}h^4\epsilon^2 m_\pi$. In the case of $\Lambda^+_c(2595)$, the lower bound of $\delta$ for the resonance to remain narrow is around merely a couple of MeVs.

Because it requires that $\Psi$ be coupled to $\pi \Sigma_c$ through the time derivative on the pion field, chiral symmetry plays an instrumental role in suppressing $k_I$. Without its constraint, we would have resorted to, on top of $(m_\pi^\star - m_\pi)/m_\pi \to 0$, a second fine-tuning on the coupling of $\Psi$ to the pion-baryon continuum, as done in Refs.~\cite{Bertulani:2002sz, Gelman:2009be}. Reference~\cite{Bedaque:2003wa} devised a power counting to describe narrow resonances with a single fine-tuning, but it applies only when the real part of the pole position $k_R$ is one order larger than $k_I$ and it requires a loose correlation between the coupling and mass splitting of the resonance field.

However, chiral symmetry facilitates a narrow resonance to be near threshold in the $S$ wave only to an extent: when $\Delta \ll \sqrt{4\pi} f_\pi$. This is an insight obtained by accounting for the fact that the pion is nonrelativistic. When $\Delta \gtrsim \sqrt{4\pi} f_\pi$ (but still within the validity range of ChPT), $k_I$ is more likely naturally sized and other decay channels are more favored than two-body interactions to generate a near-threshold resonance.

In fact, a more common mechanism to generate a near-threshold $S$-wave resonance is a two-body $S$-wave bound state weakly coupled to other decay channels. For instance, in many of its theoretical descriptions $X(3872)$ is constructed as a bound state of $D^0 \bar{D}^{*0} + \bar{D}^0 D^{*0}$ and it decays into, among others, $D^0 \bar{D}^0 \pi$~\cite{Fleming:2007rp, Hanhart:2010wh, Guo:2013sya, Jansen:2013cba, Alhakami:2015uea, Baru:2015nea}. In the particular case of $\Lambda_c^+(2595)$, the role of three-body decay into $\pi \pi \Lambda_c^+$ has been noted in Ref.~\cite{Romanets:2012ce}. While the construction in the present paper does not rule out this possibility, for when $\delta < 0$ the excited baryon indeed corresponds to a bound state, it suggests that by a small tweak of $\delta$, a narrow near-threshold two-body resonance is equally possible.

\begin{figure}
        \includegraphics[scale=0.4]{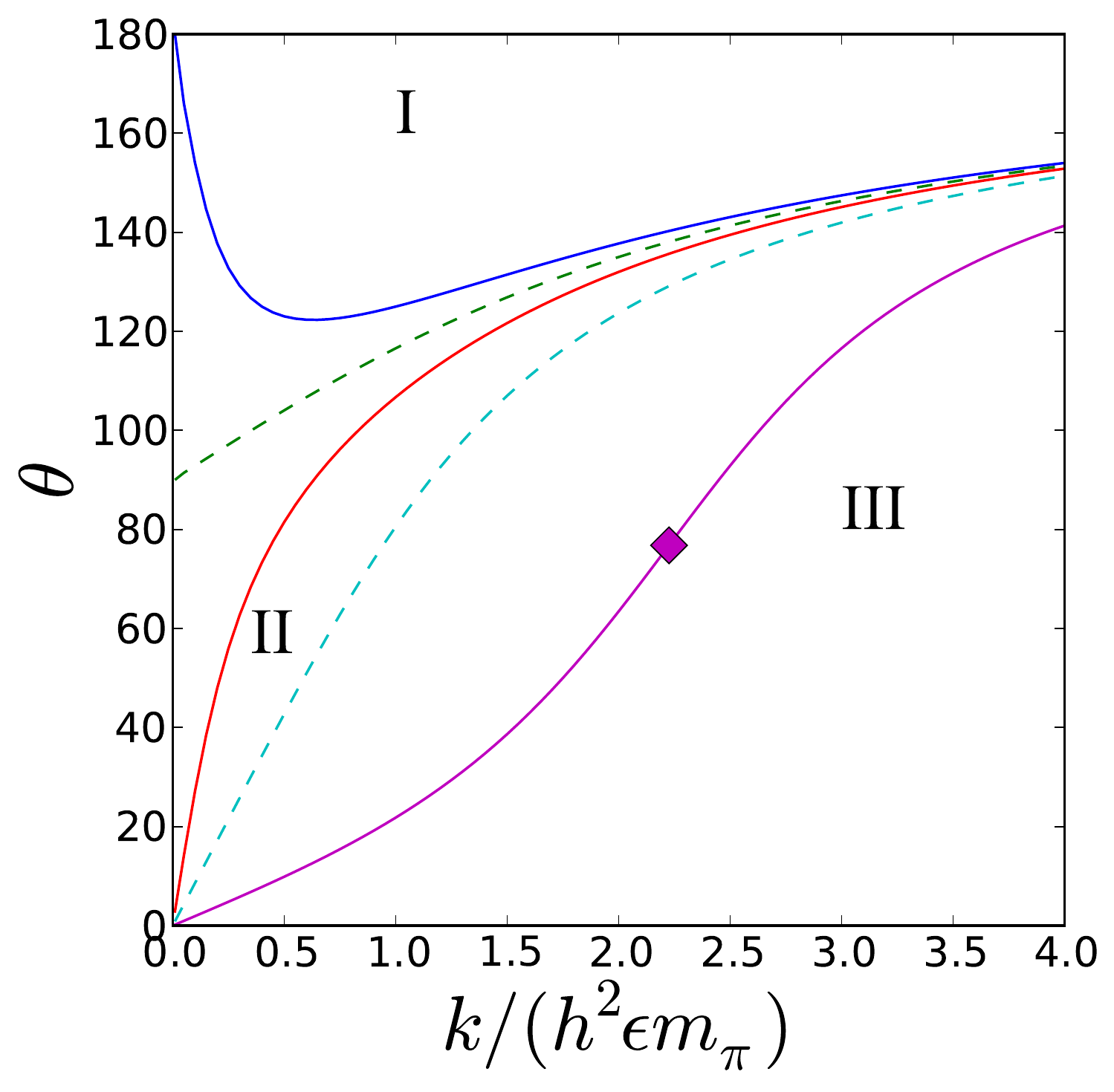}
    \caption{The LO phase shifts as functions of  $k/(h^2 \epsilon m_\pi)$, with various values of $\tilde{\delta}$. From the top down, the solid lines are the phase shifts plotted with $\tilde{\delta} = -0.2$, $0.2$, and $3$, respectively. The inflection point on $\tilde{\delta} = 3$ is marked out with a diamond. The dashed lines separate the three different regions defined in the text: the boundary between I and II is the phase shift with $\tilde{\delta} = 0$ and the one between II and III with $\tilde{\delta} = 2/3$.}
    \label{fig:theta0}
\end{figure}

Since the phase shifts can be obtained from lattice QCD via L\"uscher's formula~\cite{Luscher:1990ux}, we find it useful that the transition from the bound state to the resonance can in fact be presented by the morphing of the profile of the phase shifts. The phase shift $\theta$ at LO is most easily expressed as a function of $k/(h^2 \epsilon m_\pi)$, with $\tilde{\delta} \equiv \delta/(h^4 \epsilon^2 m_\pi)$ being the only free parameter. Shown in Fig.~\ref{fig:theta0} are the LO phase shifts, plotted with various $\tilde{\delta}$. The curves can be cast into three categories according to their geometric properties, with each category taking up a specific region.

In region I, $\delta < 0$ and a shallow bound state emerges. The phase shift function is convex over the whole domain of $k$; it descends from $180^\circ$ at threshold, as required by Levinson's theorem, turning around to start rising at the stationary point: $k_\text{stat} = \sqrt{-2 m_\pi \delta}$.

A few words about the binding energy are in order, for it is more directly linked to lattice calculations than the phase shifts. Its value around $m_\pi^\star$ at LO is found to be
\begin{equation}
  B_0(\delta; m_\pi) = \frac{h^4}{2} \epsilon^2 m_\pi  \left(\sqrt{1 - \frac{2\delta}{h^4 \epsilon^2 m_\pi}} - 1\right)^2 \, .
  \label{eqn:B0}
\end{equation}
When $\delta = 0$, the binding energy vanishes and the scattering length diverges. The universality relation, $B \propto \delta^2$ (see, for example, Ref.~\cite{Hyodo:2014bda}), is recovered for $\delta/(\epsilon^2 m_\pi) \to 0^-$,
\begin{equation}
 B =  \frac{\delta^2}{h^4 \epsilon^2 m_\pi} \left[1 +  \mathcal{O}\left(\frac{\delta}{h^2 \epsilon^2 m_\pi} \right) \right] \, .
\end{equation}
With the assumption $h = \mathcal{O}(1)$, an important revelation here is that the validity scope of universality is extremely small if $\Delta \ll \sqrt{4\pi} f_\pi$,
\begin{equation}
  \bigg{|} \frac{m_\pi - m_\pi^\star}{m_\pi} \bigg{|} \ll \left[\epsilon(\Delta) \right]^2 = \left( \frac{\Delta}{328 \text{MeV}} \right)^4 \, .
\end{equation}

The surprisingly small validity range of universality has everything to do with the emergence of a second large length scale: the effective range. We need to note that considerations of universality alone cannot capture the significance of $f_\pi$, the mass scale intimately related to chiral symmetry and its spontaneous breaking.

In region II where $\tilde{\delta}$ turns positive but is still smaller than $2/3$, the phase shift is a concave function of $k$, and has no stationary point. This region covers the $m_\pi$ gap identified by the coexistence of two virtue poles, but does not exactly coincide with it. This shows from one aspect the slight ambiguity of defining the emergence of $S$-wave resonances.

Finally, when $\tilde{\delta} > 2/3$, the phase shift functions occupy region III. They all consist of a convex segment near threshold, before becoming concave toward higher energies. The inflection point is at the origin for $\tilde{\delta} = 2/3$, but more generally its position does not have a closed form as a function of $\tilde{\delta}$. For illustration purposes, the inflection point is marked out in Fig.~\ref{fig:theta0} on the curve with $\tilde{\delta} = 3$.

To know better the uncertainty of the LO calculations and how reliable the conclusions thus drawn are, we compute the subleading corrections to the scattering amplitude. They are partly driven by the WT term, which brings no free parameters more than $h/f_\pi$, $\delta$, and $m_\pi$. Other next-to-leading order (NLO) contributions include the recoil effects of the pion. The sum of these NLO contributions can too be cast into the form of the effective range expansion,
\begin{equation}
  T^{(1)}(k) = - \frac{-\gamma_1 + \frac{r_1}{2} k^2 + P k^4 }{\left(-\gamma_0 + \frac{r_0}{2}k^2 - ik\right)^2} \, ,
\end{equation}
with
\begin{align}
    \gamma_1 &= \frac{3}{h^4} \frac{\delta}{m_\pi} \frac{\delta}{\epsilon} \, , \\
    \frac{r_1}{2} &= - \left[\frac{\delta}{m_\pi} \left(1 - \frac{3}{h^2}\right) + 2 \frac{\epsilon}{4\pi} \right] \left(h^2 \epsilon m_\pi\right)^{-1} \, , \\
    P &= \epsilon^2 h^2 \left(h^2 - \frac{3}{4}\right) \left(h^2 \epsilon m_\pi\right)^{-3} \, ,
\end{align}
where $P$ is the shape parameter, and $\gamma_1$ and $r_1$ are the corrections to the inverse scattering length and the effective range. $\epsilon/4\pi = (m_\pi/4\pi f_\pi)^2$ is the more usual ChPT expansion parameter for relativistic pions and it reflects here the recoil corrections of the pion. Compared with LO, these subleading corrections are of $\mathcal{O}(\delta/m_\pi, \epsilon^2, \epsilon/4\pi)$.

The NLO correction to the binding energy has a closed form,
\begin{equation}
  \begin{split}
    B_1(\delta; m_\pi) = \frac{h^4 \epsilon^2 m_\pi}{\sqrt{1 - 2\tilde{\delta}}}
     \left[ \epsilon^2 f(\tilde{\delta}) -\frac{\epsilon}{4\pi} \kappa^3(\tilde{\delta})  \right]    \, , \label{eqn:B1}
  \end{split}
\end{equation}
where
\begin{equation}
  \kappa(\tilde{\delta}) \equiv \sqrt{1 - 2\tilde{\delta}} - 1 \, ,
\end{equation}
and
\begin{equation}
    \begin{split}
      f(\tilde{\delta}) &\equiv h^2 \left\{ 3 h^2 \tilde{\delta}^3\, \kappa(\tilde{\delta})  - \frac{1}{2} (h^2 - 3) \tilde{\delta}\, \kappa^3(\tilde{\delta}) \right. \\
      & \quad \qquad \left. - \left(h^2 - \frac{3}{4}\right) \kappa^5 (\tilde{\delta}) \right\} \, .
    \end{split}
\end{equation}
When $\delta = 0$, $B_1 = 0$; therefore, $\delta = m_\pi^\star - m_\pi$ still holds true at NLO.

To summarize, I have explored in an $S$-wave pion-baryon system the consequences of chiral symmetry for $m_\pi$ around its critical value $m_\pi^\star$, at which point the excited baryon can be viewed as a zero-energy pion-baryon bound state. A physical realization of such systems is pion-charmed baryon $\pi \Sigma_c$, of which the resonance $\Lambda^+(2595)$ is near threshold.

Chiral symmetry is crucial in constructing a narrow $S$-wave resonance so near threshold. The nonrelativistic nature of the pion brings about one more insight: Such a near-threshold resonance is more likely when the mass splitting is smaller than an intermediate scale: $m_\pi \simeq \Delta \ll \sqrt{4\pi} f_\pi$.

Regarding pion-baryon elastic scattering, it was found that a single fine-tuning of $m_\pi^\star - m_\pi \to 0$ gives rise simultaneously to large values for both the scattering length $a$ and effective range $r$,
\begin{equation}
  a \propto \frac{\epsilon}{\delta} \, , \quad r \propto \frac{1}{\epsilon m_\pi} \, , \label{eqn:ar_con}
\end{equation}
where $\delta \equiv \Delta - m_\pi$ was found to be just $m_\pi^\star - m_\pi$, at least up to NLO. The large size of $r$ impacts the threshold physics even after the resonance collapses and becomes stable. For instance, $r$ limits considerably the applicability of universality relations based solely on a large $a$. To demonstrate, the binding energy was shown to recover the universal dependence on $\delta$, $B \propto \delta^2$, only in a tiny window $\delta/m_\pi \ll \epsilon^2$. In the region the manuscript is more concerned with, $\delta/m_\pi \ll 1$, $B$ is a more complicated function of $\delta$, shown in Eqs.~\eqref{eqn:B0} and \eqref{eqn:B1}.

Barring any further fine-tunings in higher partial waves, the emergence of two large length scales is exclusive to the $S$ wave. In higher waves chiral symmetry appears to allow only the scattering length to be fine-tuned by $m_\pi^\star - m_\pi \to 0$. For instance, the $P$-wave scattering length and effective range can be shown to scale as
\begin{equation}
  a_{P} \propto -\frac{1}{4\pi f_\pi^2\, \delta} \, , \quad r_{P} \propto -\frac{4\pi f_\pi^2}{m_\pi} \, ,
\end{equation}
where $r_{P}$ is naturally sized even when $\delta \to 0$.

\acknowledgments

I thank Han-Qing Zheng, Michael Birse, and Daniel Phillips for useful discussions, and the Institut de Physique Nucl\'eaire d'Orsay and the Espace de Structure Nucl\'eaire Th\'eorique (ESNT) at CEA Saclay for the hospitality when parts of the work were carried out. This work was supported in part by the National Natural Science Foundation of China (NSFC) under Grant No. 11375120.

\end{document}